\documentclass[10pt,conference]{IEEEtran}
\usepackage[utf8]{inputenc}

\usepackage{soul}
\usepackage{xcolor}
\usepackage{xspace}
\usepackage[hyphens]{url}
\usepackage{graphicx}
\usepackage{listings,multicol}
\usepackage[caption=false]{subfig}  
\usepackage[export]{adjustbox}
\usepackage{amsmath}
\usepackage{amssymb}
\usepackage{textcomp}
\usepackage{listings}
\usepackage{algorithm}
\usepackage[noend]{algpseudocode}
\usepackage{hyperref}
\usepackage{breakurl} 
\usepackage{cleveref}
\usepackage{booktabs}
\usepackage{multirow}

\usepackage{balance}

\makeatletter
\newcommand{\linebreakand}{%
  \end{@IEEEauthorhalign}
  \hfill\mbox{}\par
  \mbox{}\hfill\begin{@IEEEauthorhalign}
}
\makeatother

\title{Kotlin ML Pack: Technical Report}
\author{
\IEEEauthorblockN{Sergey Titov*} 
\IEEEauthorblockA{
    \textit{JetBrains Research}\\
    Paphos, Cyprus \\
    sergey.titov@jetbrains.com    
}
\and
\IEEEauthorblockN{Mikhail Evtikhiev*} 
\IEEEauthorblockA{
    \textit{JetBrains Research}\\
    Paphos, Cyprus \\
    mikhail.evtikhiev@jetbrains.com    
}
\and
\IEEEauthorblockN{Anton Shapkin*} 
\IEEEauthorblockA{
    \textit{JetBrains Research}\\
    Paphos, Cyprus \\
    anton.shapkin@jetbrains.com    
}
\linebreakand
\IEEEauthorblockN{Oleg Smirnov} 
\IEEEauthorblockA{
    \textit{JetBrains Research}\\
    Amsterdam, the Netherlands \\
    oleg.smirnov@jetbrains.com    
}
\and
\IEEEauthorblockN{Sergei Boytsov} 
\IEEEauthorblockA{
    \textit{JetBrains Research}\\
    Berlin, Germany \\
    sergey.boytsov@jetbrains.com
}
\and
\IEEEauthorblockN{Dariia Karaeva} 
\IEEEauthorblockA{
    \textit{JetBrains Research}\\
    Limassol, Cyprus \\
    dariia.karaeva@jetbrains.com
}
\linebreakand
\IEEEauthorblockN{Maksim Sheptyakov} 
\IEEEauthorblockA{
    \textit{JetBrains Research}\\
    London, UK \\
    maksim.sheptyakov@jetbrains.com
}
\and
\IEEEauthorblockN{Mikhail Arkhipov} 
\IEEEauthorblockA{
    \textit{JetBrains Research}\\
    Amsterdam, the Netherlands \\
    mikhail.arkhipov@jetbrains.com    
}
\and
\IEEEauthorblockN{Timofey Bryksin} 
\IEEEauthorblockA{
    \textit{JetBrains Research}\\
    Limassol, Cyprus \\
    timofey.bryksin@jetbrains.com    
}
\linebreakand
\IEEEauthorblockN{Egor Bogomolov} 
\IEEEauthorblockA{
    \textit{JetBrains Research}\\
    Amsterdam, the Netherlands \\
    egor.bogomolov@jetbrains.com
}
}

\begin{document}

\maketitle

\begin{abstract}
In this technical report, we present three novel datasets of Kotlin code: KStack, KStack-clean, and KExercises. 
We also describe the results of fine-tuning CodeLlama and DeepSeek models on this data. 
Additionally, we present a version of the HumanEval benchmark rewritten by human experts into Kotlin --- both the solutions and the tests. 
Our results demonstrate that small, high-quality datasets (KStack-clean and KExercises) can significantly improve model performance on code generation tasks, achieving up to a 16-point increase in pass rate on the HumanEval benchmark. 
Lastly, we discuss potential future work in the field of improving language modeling for Kotlin, including the use of static analysis tools in the learning process and the introduction of more intricate and realistic benchmarks.
\end{abstract}

\section{Introduction}\label{sec:introduction}

\renewcommand*{\thefootnote}{\fnsymbol{footnote}}
\footnotetext[1]{The ﬁrst three authors contributed equally to this work.}
\renewcommand*{\thefootnote}{\arabic{footnote}}

To stay relevant in the world of AI revolution, the language needs to be well-represented in the machine learning community. 
The less a language is represented, the fewer works are done around it. 
This leads to a lower quality of generated code, which in turn results in the decreased usage of the language and even poorer representation.
To avoid such a situation for Kotlin, we propose the \textit{Kotlin ML Pack}. 
The goal of the project is to provide all the necessary tools, data, and models to promote code modeling tasks for Kotlin.

In this report, we present three Kotlin datasets: 
\begin{itemize}
    \item KStack\footnote{KStack: \href{https://huggingface.co/datasets/JetBrains/KStack}{https://huggingface.co/datasets/JetBrains/KStack}} — the biggest collection of permissively licensed Kotlin files;
    \item KStack-clean\footnote{KStack-clean: \href{https://huggingface.co/datasets/JetBrains/KStack-clean}{https://huggingface.co/datasets/JetBrains/KStack-clean}} — a highly filtered version of KStack containing twenty-five thousand high-quality examples;
    \item KExercises\footnote{KExercises: \href{https://huggingface.co/datasets/JetBrains/KExercises}{https://huggingface.co/datasets/JetBrains/KExercises}} — translated and improved version of the Python exercises dataset~\cite{pexer}. 
\end{itemize}

More information about dataset sizes, filtration techniques, and comparison with other datasets can be found in Section~\ref{sec:data}.

We also provide a version of HumanEval~\cite{chen2021codex} translated into Kotlin by human experts. 
In Section ~\ref{sec:evaluation}, we demonstrate problems with the current evaluations of code generation for Kotlin and elaborate on the translation process. 

Lastly, we provide multiple checkpoints for fine-tuned models. 
We took two base models: CodeLlama \cite{roziere2023code} and Deepseek-coder \cite{guo2024deepseek}, and for each of them, performed fine-tuning on variations of the KStack dataset and KExercises. 
More information on the fine-tuning process, techniques used during the training, and results can be found in Section~\ref{sec:modeling}.

We believe that this work provides the bedrock for further research on Kotlin code generation. 
In Section~\ref{sec:future-work}, we describe possible directions for future research, including such promising ideas as using a compiler during the training process and designing a more realistic benchmark for Kotlin. 

\begin{figure*}[ht]
    \centering
    \includegraphics[width=\textwidth]{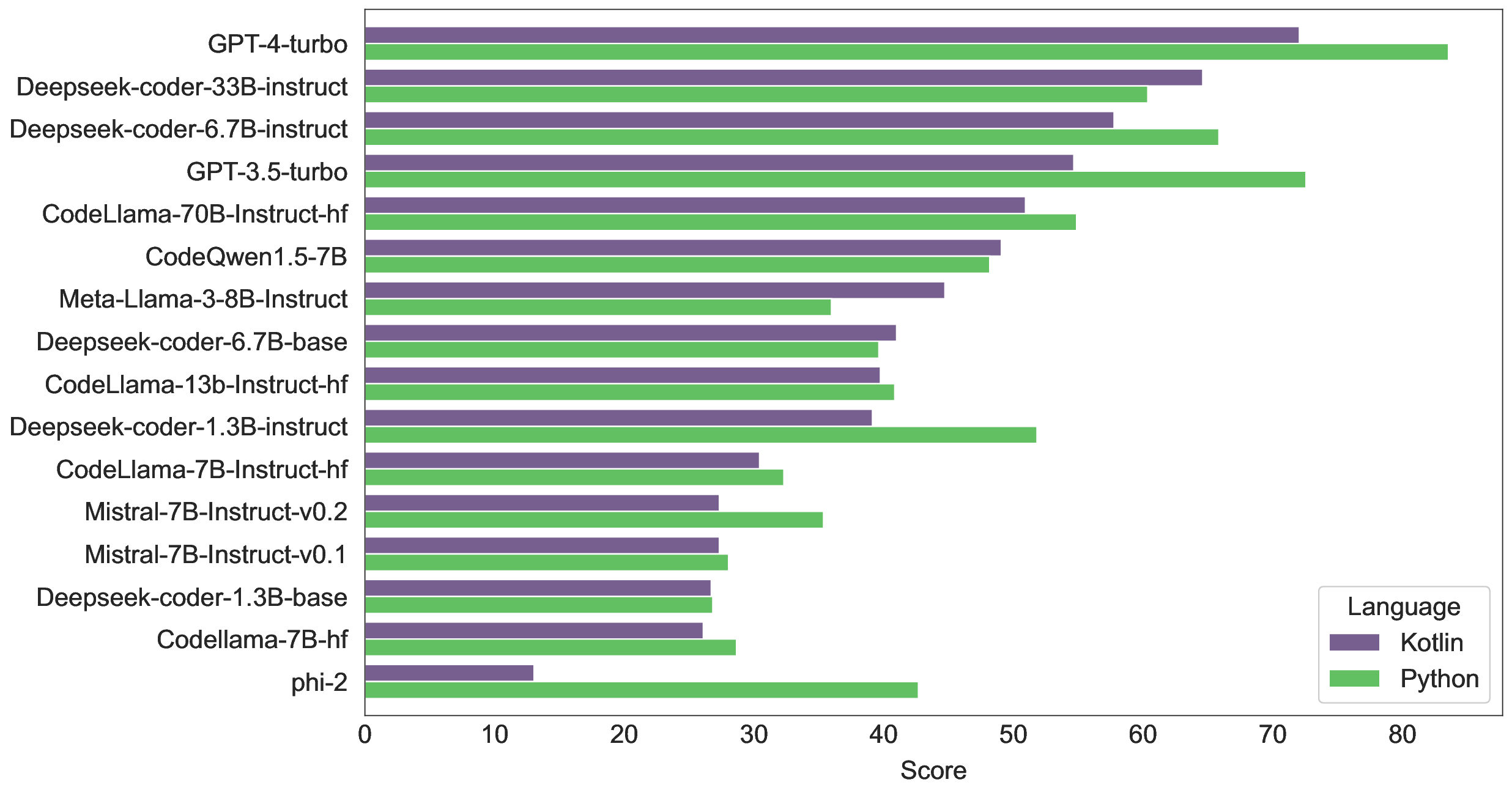}
    \vspace{-0.3cm}
    \caption{Kotlin and Python HumanEval Scores for various models.}\vspace{-0.3cm}
    \label{fig:humaneval_sota}
\end{figure*}
\section{Current state of the art \\ of Kotlin code generation}\label{sec:background}

First of all, we needed to determine how well the current generation of models can generate Kotlin code. 
To benchmark Kotlin code generation, we used our human-translated version of the HumanEval benchmark, specifically adapted for Kotlin. 
More details on the setup of the Kotlin HumanEval are provided in Section~\ref{sec:evaluation}. 
Additionally, we test each model on the Python HumanEval to control the used methodology and to compare the model performance between different programming languages. The findings from these benchmarks are shown in~\Cref{fig:humaneval_sota}.

Interestingly, the performance gap between larger models, such as CodeLlama-70B, DeepSeek-coder-33B, and the GPT family, is not as significant as the difference in their number of parameters. 
This is in contrast to smaller models, like the 7B models, where the performance gap is more pronounced.
Improvements in performance thus do not appear to depend linearly on the parameters number. 

Importantly, in most cases, the performance on Kotlin is worse than the performance of these models on Python. 
One of the reasons for this is the popularity of Python in the research community, which impacts both model training and evaluation. 
Many models are evaluated on tasks involving Python code, and their performance on Kotlin is not considered. 
To achieve better quality in Python tasks, models are tuned on Python-specific data. 
Furthermore, we observe that models fine-tuned on instruction data for Python exhibit improved performance on the HumanEval benchmark for both languages. 

Despite the development of specialized coding models, OpenAI's models continue to set the state-of-the-art standard in coding tasks. 
This could be attributed not only to the size of the models but also to the fact that natural language comprehension can play a crucial role in code generation from natural language. 
Training on natural language texts provides contextual understanding that enhances the model's ability to interpret and generate code more effectively.

As a result, we can conclude that the current generation of models underperforms when solving coding tasks in Kotlin. 
While large proprietary models provide good scores on the tasks, there is significant space for improvement for open source models.

\section{Kotlin data}\label{sec:data}

Data is the cornerstone of machine learning in every domain. 
Python appears to be one of the most represented programming languages in machine learning --- not only is Python the third language by popularity in the biggest code dataset --- the Stack~\cite{Kocetkov2022TheStack}, --- there are also multiple high-quality and domain-specific datasets available for it. 
For example, there are datasets like APPS~\cite{hendrycksapps2021} and Python exercises~\cite{pexer}, which contain high-quality Python tasks with solutions, as well as the DS-1000 dataset~\cite{lai2023ds}, which focuses on data science-specific tasks for Python.

These datasets were created to teach and measure various aspects of Python language modelling. Following this, we focus on two main datasets for Kotlin: the language corpus and the instruction dataset. 

\subsection{KStack: Kotlin language corpus}

A large language corpus is needed to fine-tune various models on Kotlin.
Even though some popular datasets, such as the Stack~\cite{Kocetkov2022TheStack}, already have Kotlin code, we present the most complete permissively licensed up-to-date collection of open-source Kotlin code.

We collect repositories from GitHub where the main language is Kotlin, as well as repositories containing Kotlin files that have ten or more stars (as of February 2024).
Additionally, we gather repositories with Kotlin files from Stack v1.2~\cite{Kocetkov2022TheStack}. 
We identify Kotlin files using \textit{go-enry}~\cite{goenry}, and include files with extensions such as \textit{.kt}, \textit{.kts}, and \textit{.gradle.kts}. 
Next, we use the file content hash to conduct full deduplication, as well as near deduplication using the same method as in Stack v1.2~\cite{Kocetkov2022TheStack}.
We then leave a single representative from each near-deduplicated cluster, picking the file from the repository with the most stars.
We also filter the dataset, leaving only repositories with permissive licenses.
We use the license information provided by GitHub, and fall back to \textit{go-license-detector}~\cite{golicense} if license information is not available on GitHub. 
The list of permissive licenses used in KStack can be found at our HuggingFace page.\footnote{List of licenses in KStack: \url{https://huggingface.co/datasets/JetBrains/KStack/blob/main/licenses.json}}
Finally, we filter personal and sensitive information using the \textit{star-pii} model~\cite{lozhkov2024starcoder}.

\subsection{KStack-clean: Learning the code quality}

After collecting the almost full corpus of Kotlin code in KStack, we continue with creating a dataset of specifically high-quality Kotlin code. 
Using curated datasets to fine-tune a model can provide larger improvements than fine-tuning it on a bigger corpus of non-curated data~\cite{gunasekar2023textbooks}.

To find high-quality data in our dataset, we build a classifier that predicts the quality of the code. 
We start by labeling a small portion of KStack documents (128K examples) with LLMs (Mistral-7B-Instruct-v0.2 and GPT-3.5-Turbo~\cite{chatgpt}) in a zero-shot setting. 
We used pairwise strategy for labeling. 
In this strategy, we score a pair of files by asking which file has greater ``educational value for learning algorithms in Kotlin''. We prepend tokens \texttt{A} and \texttt{B} before each file. To get the score for the file, we calculate the difference between log probabilities of \texttt{A} and \texttt{B} tokens.
To avoid the effects of ordering in the prompt, we use the following formula: 
    $$s(f) = \frac{\bigl(s(f, c)_A - s(f, c)_B\bigr) + \bigl(s(c, f)_B - s(c, f)_A\bigr)}{2}$$
where $f$ is the file, for which we are estimating quality, $c$ is the highest rated sample, $s(f, c)_A$ is the probability of choosing the code labeled as \texttt{A} (first argument $f$) over the code labeled as \texttt{B} (second argument $c$).

While scoring each pair is a computationally expensive process, we designed an approximation in a form of three-pass procedure. 
We begin by scoring a random sample against the full dataset to get a relative quality score for each example.
Next, we selected the example with the highest score and repeated the process, scoring it against the full dataset to get a second set of scores. 
Finally, we repeated the procedure with the highest score sample from the second pass. 
To obtain the final scores, we averaged the scores from second and third passes (omitting the random sample scores) for the 128 thousand examples.

After we get the scores for the initial sample of 128K examples, we train a binary classifier based on 220M-CodeT5+ \cite{wang2023codet5plus}. 
Using the obtained scores, we mark the top 5\% of samples as positive examples (high-quality examples), and the rest we mark as negatives. 
The classifier is trained to distinguish these examples for three epochs and then applied to the entire KStack dataset. 
For each sample in the dataset, the log-probability of the positive class is used as a quality estimate. Finally, to obtain KStack-clean dataset, we select 25,000 samples from KStack with the highest probabilities of the positive class.

\subsection{Comparison with other datasets}

As a result of our data collection procedure, we end up with two datasets: \textit{KStack} and \textit{KStack-clean}. In Table~\ref{table:dataset_comp}, we compare the descriptive statistics for these two datasets and the Stack v2~\cite{lozhkov2024starcoder}. 

\begin{table}[t]
\centering
\resizebox{\columnwidth}{!}{%
\begin{tabular}{@{}lcccc@{}}
\toprule
                      & \multicolumn{1}{l}{\textbf{Files}} & \multicolumn{1}{l}{\textbf{Repositories}} & \multicolumn{1}{l}{\textbf{Lines}} & \multicolumn{1}{l}{\textbf{Tokens}} \\ \midrule
\textbf{The Stack v2}          & 2M                        & 109,547                           & 162M                      & 1.7B                       \\
\textbf{KStack}       & 4M                        & 163,310                           & 293M                      & 3.1B                       \\
\textbf{KStack-clean} & 25,000                     & 3,366                             & 2M                        & 22M   \\\bottomrule
\end{tabular}
}

\vspace{0.3cm}
\caption{The descriptive statistics for the collected datasets. To calculate token numbers we use CodeLlama tokenizer.}\vspace{-0.5cm}
\label{table:dataset_comp}
\end{table}

\subsection{KExercises: Kotlin instructions dataset}

Our next goal with regards to the dataset development is  creation of the Kotlin instructions dataset. 
Typically, such datasets consist of sets of instructions or tasks along with their solutions. 
Training on such data aids models in better comprehending the relationship between natural and programming languages. 

There is a number of such datasets available for the Python programming language, like APPS~\cite{hendrycksapps2021} and Python Code Exercises~\cite{pexer}. 
Moreover, there are also multi-language datasets, such as CodeAlpaca~\cite{codealpaca} and CommitPackFT~\cite{muennighoff2023octopack}. 
However, these datasets only have a relatively modest representation of Kotlin code or do not contain Kotlin at all. 

Kotlin could be considered a low-resource language due to the scarcity of publicly available data and the limited opportunities for improvement using data collected from open-source projects.
This motivated us to use synthetic data for instruction tuning. 
Generating synthetic data can present many challenges, and one of the biggest of them is a low sample diversity. 
According to Gunasekar et al.~\cite{gunasekar2023textbooks}, generated data should encompass a broad spectrum of coding concepts, skills, and scenarios, varying in difficulty, complexity, and style. 
Since there are many synthetic Python datasets that have addressed these potential issues, we decided to adapt one of these datasets by translating it from Python to Kotlin, rather than creating an entire dataset from scratch.

For this purpose, we used the CodeExercises dataset~\cite{pexer} that replicates the steps outlined in the work of Gunasekar et al.~\cite{gunasekar2023textbooks}, demonstrating its high quality and effectiveness.
We used GPT-3.5-turbo to translate the data from Python to Kotlin (see the translation prompt in~\Cref{figure:p2k_translation}).
We iteratively translated segments of data and monitored the downstream Kotlin generation quality during validation. 
Additionally, after the translation, we manually reviewed a sample of the data to ensure the accuracy of the translations. 
Finally, we compiled an instruction dataset comprising 15K Kotlin tasks (approximately 3.5 million tokens, 335K lines of code). 
A more detailed investigation of the impact of the size of the synthetic sample on the final quality of the fine-tuned models is left for future work.

\begin{figure}[H]
\begin{verbatim}
System: You are a helpful assistant.
User: Rewrite to Kotlin (do not forget 
about docstring):\n\nPYTHON_CODE
\end{verbatim}
\caption{Prompt for Python to Kotlin translation.}
\label{figure:p2k_translation}
\end{figure}

\section{Kotlin evaluation}\label{sec:evaluation}

A vital aspect of machine learning is accurate and efficient evaluation procedures, with HumanEval being a standard for code LLMs evaluations~\cite{chowdhery2023palm,wei2022chain,li2023starcoder,lozhkov2024starcoder}. 
Though initially designed for the Python programming language, since its debut in 2020, HumanEval has been translated into multiple programming languages. 
It has also been adapted for use with compiled languages and has been expanded with new tasks.

\subsection{HumanEval for Kotlin}
While there is a version of HumanEval for Kotlin~\cite{mbxp_athiwaratkun2022}, it required significant improvement before using.
Most of the issues we discovered were related to faulty prompts.
The most common prompt issue was too generic variable type in the Kotlin function signature.
In such cases, it is impossible to apply many built-in Kotlin methods (\textit{e.g.}, one cannot use \textit{sort} method on Kotlin array of \textit{Any} type).
The other source of issues were tests.
The difference in rounding procedures between Kotlin and Python caused tests to fail on a perfectly reasonable Kotlin program.
For example, for \textit{HumanEval task \#2}, the task was: 
\begin{verbatim}
Given a positive floating point number,
it can be decomposed into integer part 
(largest integer smaller than given number)
and decimals (leftover part always 
smaller than 1).
Return the decimal part of the number.
\end{verbatim}
For this case, the following solution
\begin{verbatim}
fun truncate(number : Double) : Double {
    return number - Math.floor(number)
}
\end{verbatim}
fails due to the rounding error of less than $1e-8$.

To solve the problems with the existing Kotlin HumanEval, we asked human experts to rewrite the HumanEval from scratch. 
All HumanEval solutions and tests in Kotlin were thus written by an expert competitive programmer with six years of experience in Kotlin, and independently reviewed by a programmer with four years of experience in Kotlin.
The tests we implemented are equivalent to the original HumanEval tests for Python, and we fixed the prompt signatures to address the generic variable signature we described above.

\subsection{Evaluation setup for code generation}

To evaluate the code generation quality for our fine-tuned models, we use our version of HumanEval and the greedy generation strategy.
The generation worked as follows.

To prompt the model, we use the prompt from our version of Kotlin HumanEval (\textit{You are an expert Kotlin programmer...}). 
We use the \textit{generate} method from HuggingFace \textit{transformers}, greedy generation, and force the model to generate from 128 to 256 new tokens. 
To speed up evaluation and simplify answer parsing, we use the early stopping token sequence ``\textbackslash n\}\textbackslash n'', which corresponds to the end of the Kotlin method according to the Kotlin coding conventions.
Additionally, we remove all comments from the answer as these do not affect the performance and can clutter the outputs when one wants to read them. 
Since some models repeat part of the prompt in the answer (which we do not deem to be a model failure), we find the first definition of the function in the answer (the line starts with ``\textit{fun}''), and delete this line and all the lines before from the answer before concatenating it to the prompt.

We then pass the resulting code to the MXEval \cite{mbxp_athiwaratkun2022} to score compilation and pass rates. 
We report the following metrics for each model:
\begin{itemize}
    \item \textbf{Pass@1} — percentage of generations that successfully passed the tests;
    \item \textbf{Test Error Rate} --- percentage of generations that failed one or more tests;
    \item \textbf{Compilation Error Rate} — percentage of generations failed on the compilation step;
    \item \textbf{Out of Time Error Rate} --- percentage of methods that failed to pass one or more tests in 15 seconds.
    \item \textbf{Runtime Error rate} — percentage of generations failed during runtime. 
    In addition to the usual runtime errors, these include methods which have \textit{TODO()} in the body.
\end{itemize}

\textbf{Pass and error rates as a measure.} 
Typical usage of a test-based benchmark involves reporting only the pass rate as a measure of the model's success. 
The pass rate is a reasonable proxy for assessing code generation from description on simple tasks. 
In addition to the pass rate, we suggest using \textbf{syntax error rate} as a complementary metric that tracks how the model comprehends Kotlin.
The syntax error rate is a sum of runtime error rate and compilation error rate.
Together with the pass rate, this metric helps to distinguish between the model's language knowledge and its level of task understanding.
On a more granular level, one can improve understanding of the model performance by examining the errors generated by the code. 
Since the size of the HumanEval benchmark allows one to go over model outputs one by one, we believe that manual checks can help to find the weak spots of the model.
This approach also enabled us to find the issues with the original HumanEval benchmark.
    
\subsection{Evaluation setup for code completion}

In addition to the evaluation of code generation, we also provide the evaluation of code completion based on a small holdout set extracted from the KStack, consisting of 630 examples. 
To score completion, we use all models in FIM setup if possible and then calculate the percentage of exact matches for the first generated line.

We believe that completion metrics help us not only measure the performance of a model for a completion task, but also control model for overfitting on the HumanEval tasks.
As the base models we use have some Kotlin capabilities from pre-training, fine-tuning may decrease their performance on the code completion tasks.
\section{Learning Kotlin}\label{sec:modeling}

To showcase our dataset, we train several models using different setups. 
The section below presents the experimental framework and the main results.

\subsection{Base models}

We use Code Llama 7B~\cite{codellamacard}, Deepseek-coder-6.7B~\cite{ds67b} and Deepseek-coder-1.3B~\cite{ds1.3b} as our base models.

Code Llama 7B~\cite{codellamacard} is an autoregressive language model that uses optimized transformer architectures.
It supports infilling text generation, fine-tuning with up to 16K tokens context sizes, and supports up to 100K tokens at inference time.

Deepseek-coder-6.7B base model, implemented by Deepseek~\cite{guo2024deepseek}, is a 6.7B-parameter model with Multi-Head Attention trained on two trillion tokens of natural language texts in English and Chinese. 
It is also pre-trained on project-level code corpus by employing a window size of 16K tokens and an extra fill-in-the-blank task to support project-level code completion and infilling. 
Deepseek-coder-1.3B shares the same architecture and training procedure, but with fewer parameters.  

\subsection{Datasets}

As described above, we used several of our datasets as part of the training setup.

\subsubsection{KStack}

To improve the dataset even more, we developed a number of filtration techniques. 
First of all, we started with filtering out files that we deem to be low quality: 
\begin{itemize}
    \item We filter out files that belong to low-popular repos (the sum of stars and forks is less than 6).
    \item Next, we filter out files that belong to the repos with less than five Kotlin files.
    \item Finally, we remove files that have less than 20 SLOC.
\end{itemize}
We then clean the remaining dataset entries by removing all non-ASCII entries, removing all package lines such as \textit{package kotlinx.coroutines.channels}, and removing half of the import lines in a probabilistic way (the line is removed with a probability of $0.5$).
This was done because we found out empirically that many Kotlin files contain many import lines. 
When one trains the model on them, the model can lock in generating import after import, when given the beginning of the file as the context.
As for the package information, it is specific to each project and, outside of the project adaptation domain, will only add noise to the training data.

\subsubsection{KStack-clean}

As this dataset already presents a highly preprocessed version of KStack, no further prepossessing was done. However, we study an alternative version of filtering which uses OpenAI GPT-3.5 as a base classifier. 

\subsubsection{KExercises}

Because of the synthetic nature of this dataset, its size is a hyperparameter and we ran multiple experiments with datasets of varying size, resulting in the optimal size of fifteen thousand examples. 

\subsection{Training setup} 

For all cases, we perform all fine-tuning on an NVidia A100 GPU in bf16 precision, using AdamW  optimizer~\cite{Loshchilov2017DecoupledWD}. We perform validation on the small holdout part of KStack consisting of 630 handpicked examples. 

All other hyperparameters are varied between different setups. For precise data on the learning rate, batch size, and learning schedule, please refer to the corresponding model card on Hugging Face. 

Additionally, we use several techniques to stabilise the training process: 
\begin{itemize}
    \item \textbf{Z-loss}: we use z-loss~\cite{chowdhery2023palm} to control possible divergence of logit outputs. 
    Let $y_j$ denote the output logits, so the class probabilities are computed as $p_j = y_j / Z$, where $Z = \sum \exp(y_j)$. 
    The instability may happen closer to the end of the training, when the logits become very negative~\cite{wortsman2023small}.
    The auxiliary z-loss term is given by $\log^2 Z$, and it forces $\log Z$ to remain close to zero. 
    \item \textbf{Weight decay}: similarly to the z-loss, weight decay can be used to control output logits divergence, as it prevents weights from growing too large.
    Since we use standard weight decay PyTorch optimization, we set the weight decay coefficient to be quite large, as it then multiplies a relatively small maximal learning rate~\cite{wortsman2023small}.
    \item \textbf{Dynamic $\beta$}: following the Palm technical report~\cite{chowdhery2023palm}, we try changing $\beta_2$ AdamW learning parameter as $\beta_2 = 1 - k^{-0.8}$, where $k$ is step number. 
    While Chowdhery et al.~\cite{chowdhery2023palm} claim this technique helps because rare embedding tokens can have poorly estimated second moments over shorter windows, in our setup it fails to improve model fine-tuning results.
    \item \textbf{Decreasing $\epsilon$ for AdamW}: we follow the suggestion of Wortsman et al.~\cite{wortsman2023small} to decrease the parameter $\epsilon$ of AdamW, since the default value of $\epsilon = 10^{-8}$ may dampen the model updates for larger models. 
    We find that in our case, the setting of $\epsilon = 10^{-16}$ slightly improves both train loss and downstream benchmark scores at no extra costs.
    \item \textbf{Gradient norm clipping}: to limit the impact of data outliers on the training process, we use gradient norm clipping for training on the KStack dataset. 
    We choose gradient clipping so that very few gradients are clipped, avoiding the effective decrease of the learning rate caused by the aggressive gradient clipping.
    \item \textbf{Warm-up}: We use longer warm-up period to reduce the model's learning rate sensitivity. 
    While using warm-up for training Transformer models with an adaptive optimizer is a de-facto standard by now, using warm-up period length of up to 10\% of the train dataset allows training models at a higher learning rate without facing instabilities~\cite{wortsman2023small}.
    We found this technique to be especially useful for the smaller datasets we use.
\end{itemize}
\subsection{Findings} 

\begin{table*}[ht!]
\centering
\resizebox{\textwidth}{!}{%
\begin{tabular}{llccccc|cc}
\toprule
\textbf{Model}                       & \multicolumn{1}{c}{\textbf{Modification}} & \textbf{Pass rate} & \textbf{Test fail rate} & \textbf{\begin{tabular}[c]{@{}c@{}}Compilation \\ error rate\end{tabular}} & \textbf{\begin{tabular}[c]{@{}c@{}}Runtime \\ error rate\end{tabular}} & \textbf{Out of time rate} & \textbf{Syntax error rate} & \textbf{\begin{tabular}[c]{@{}c@{}}Completion \\ (Exact match, first line)\end{tabular}} \\ \midrule
\multirow{4}{*}{CodeLlama-7B}        & Base                                      & 26.09              & 50.31                   & 19.25                                                                      & 3.73                                                                   & 0.62                      & 22.98                      & 0.388                                                                                    \\
                                     & KStack                                    & 29.19              & 47.2                    & 18.63                                                                      & 4.35                                                                   & 0.62                      & 22.98                      & 0.396                                                                                    \\
                                     & KStack-clean                              & 37.89              & 42.86                   & \textbf{15.53}                                                                      & 3.11                                                                   & 0.62                      & \textbf{18.64}             & \textbf{0.403}                                                                           \\
                                     & Kexer                                     & \textbf{42.24}     & \textbf{37.89}          & 17.39                                                             & \textbf{1.86}                                                          & 0.62                      & 19.25                      & 0.344                                                                                    \\ \midrule
\multirow{2}{*}{Deepseek-7B}         & Base                                      & 40.99              & 37.27                   & 11.8                                                                       & 9.32                                                                   & 0.62                      & 21.12                      & 0.403                                                                                    \\
                                     & Kexer                                     & \textbf{55.28}     & \textbf{29.19}          & \textbf{11.18}                                                             & \textbf{4.35}                                                          & \textbf{0}                & \textbf{15.53}             & \textbf{0.411}                                                                           \\ \midrule
\multirow{3}{*}{Deepseek-coder-1.3B} & Base                                      & 26.71              & 54.04                   & \textbf{14.91}                                                             & 4.35                                                                   & \textbf{0}                & 19.26                      & 0.403                                                                                    \\
                                     & KStack                                    & 27.95              & 51.55                   & 16.77                                                                      & 3.11                                                                   & 0.62                      & 19.88                      & \textbf{0.404}                                                                           \\
                                     & Kexer                                     & \textbf{36.65}     & \textbf{44.72}          & 16.15                                                                      & \textbf{2.48}                                                          & 0                         & \textbf{18.63}             & 0.388                                                                                    \\ \bottomrule \\
\end{tabular}
}
\caption{Fine-tuning results}
\label{table:results}
\end{table*}

Results for all models can be found in Table~\ref{table:results}. 
We observe improvements across all approaches that we used. 
We achieve the most significant boost with a combination of Deepseek-coder-6.7B and the fine-tuning on the KExercises dataset, resulting in the pass rate of 55.28\%.
Fine-tuning on instructions showed great results on other two base models as well. 
At the same time, fine-tuning on the full dataset shows weak results, increasing the pass rate for CodeLLama by only 3 percentage points. 
The clean version of the KStack shows much better results during fine-tuning, but the pass rate is still lower than the one that we achieved with the KExercises dataset. 
We suggest that this could be caused by a mismatch between the tuning dataset and the evaluation — the KExercises dataset allows the model to learn to follow instructions rather than learn Kotlin. 
This is supported by the completion rate data, where models tuned on KExercises dataset are outperformed by other models we fine-tune.
However, the syntax error rate data for the KExercises-tuned model shows that the fine-tuning improved Kotlin comprehension, and only completion capabilities may have been affected.
Thus, while we achieve our most significant score improvements with KExercises dataset, we believe that other datasets are still valuable for learning other tasks, such as completion or generation of documentation.

\begin{figure}[ht]
    \centering
    \includegraphics[width=0.5\textwidth]{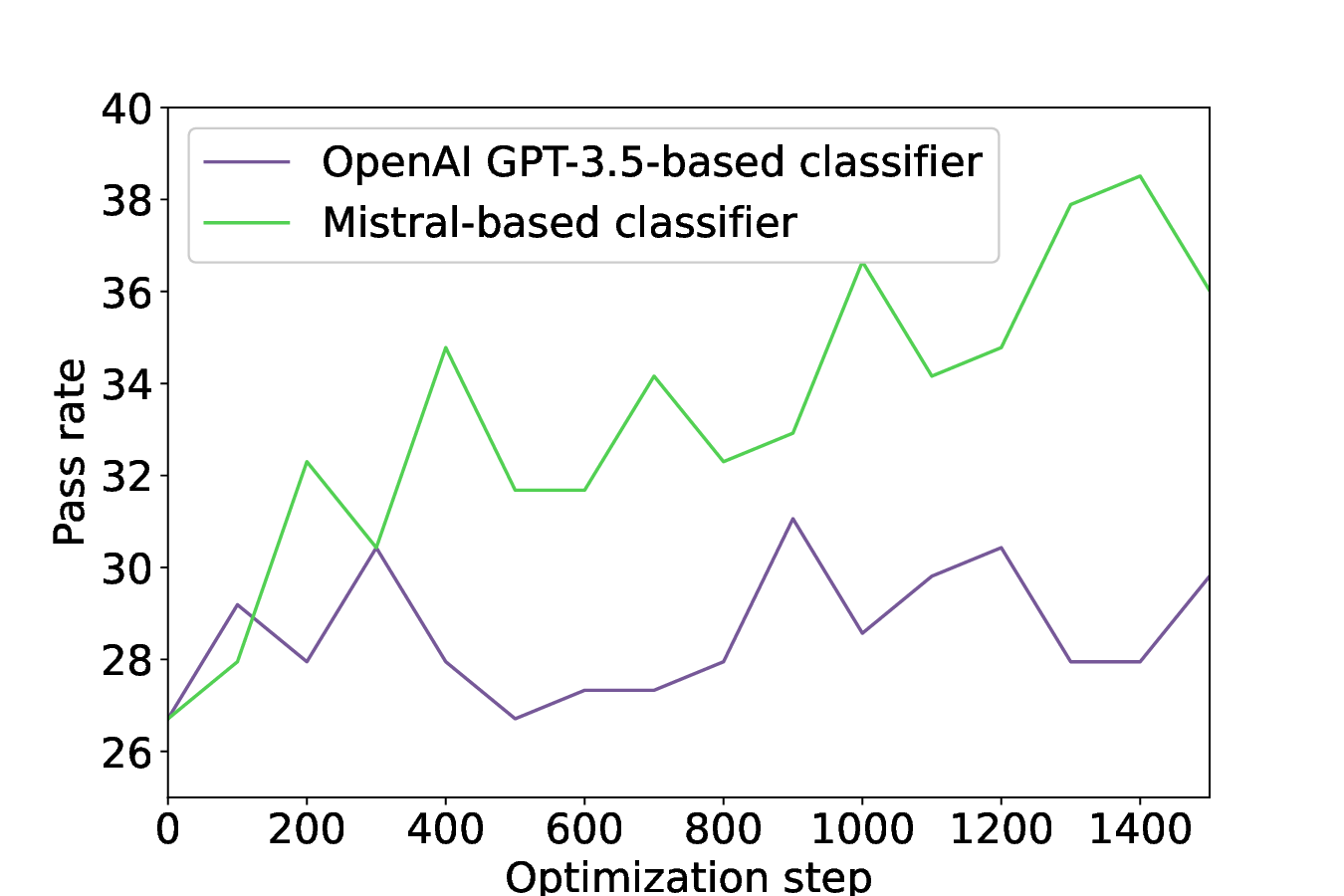}
    \vspace{-0.3cm}
    \caption{Pass rate on HumanEval for Kotlin for different filtration strategies of the \textit{KStack-clean} dataset, finetuning the CodeLlama-7B model.}\vspace{-0.3cm}
    \label{fig:kstack-fine-tune}
\end{figure}

Additionally, we performed an experiment with different base classifiers for the data cleaning procedure used in KStack-clean. 
We found that the same classifier built with OpenAI GPT-3.5 provides significantly lower quality of the sample ranking than Mistral. 
This can be clearly seen from the difference in the pass rate dynamics presented in Figure~\ref{fig:kstack-fine-tune}. 
We attribute this to the noise in the log-probabilities of the completion distribution in the OpenAI API. 
We observe high variability of log-probabilities even when the query is exactly the same. 
This noise might be artificially added to the responses as a defence against distillation.

\section{Discussion}

\subsection{Learning the language}

\subsubsection{Base models}
We used CodeLlama-7B and Deepseek-coder-6.7B as the base models for fine-tuning.
While these models are more suited for code generation in Python (see Table~\ref{table:results}), they nevertheless possess non-trivial Kotlin capabilities and were apparently trained on some amount of Kotlin code.
Therefore, training these models in some random setup on a dataset of low-quality Kotlin code can worsen their performance.
Correspondingly, the improvements in fine-tuned models caused by the dataset choice are not a low-hanging fruit of making the model learn at least something.
We consider these improvements to be a utilization of some of the unused model capabilities at a cost of one to 24 NVidia A100 GPU hours, depending on the model and dataset choice.

\subsubsection{Tracking model performance}
To assess the changes in model performance, we focus on the pass rate, and use syntax error rate and code completion as the auxiliary metrics.
The reasons for that are twofold.

First, the pass rate is a standard measure for the test-based benchmarking.
In addition to its benefits we describe in \Cref{sec:evaluation}, using pass rate as a main metric allows a simple apples-to-apples comparison of the dataset impact to the other model improvement techniques used by researchers and practitioners.
Second, HumanEval benchmark is a well-established, popularr benchmark, which we further cleaned and improved. 
Our completion benchmark, however, was not verified with the same level of rigor.

Nevertheless, we strongly believe that using just the pass rate to assess the model is not sufficient to gauge its performance.
For example, one can accidentally overfit the model on the benchmark test dataset~\cite{schaeffer2023pretraining} (which is all the more possible once the benchmarks become public), artificially inflating the pass rate.
In such a case, a completion dataset will allow to check if there was such an overfit, as the code completion capabilities of the model will be likely strongly affected.
The syntax error rate, on the other hand, allows to estimate the model's language capabilities, and thus augments the code completion score information.

\subsection{Impact of datasets on model performance}
\subsubsection{KStack}
Despite KStack being the largest dataset we used, fine-tuning models on it provides only modest improvements in the pass rate, as well as inconclusive changes in completion rate and syntax error rate. 
We surmise that these little improvements are due to the relatively low quality of the dataset, which was not amended by the filtering.
This hypothesis is supported by larger improvements on KStack-clean (which we discuss below).

\subsubsection{KStack-clean}
Even though KStack-clean dataset is circa 100 times smaller than the full KStack dataset, we observe significant pass rate improvements of over 8 percentage points for the models fine-tuned on it. 
This performance improvement is accompanied by a significant decrease in syntax error rate (15\%, or more than 3 points), and a smaller improvement of the code completion score. 
This success proves the viability of using a classifier to find the best subsample of the dataset for such widely-applicable area as algorithms, and overall shows the model improvement in Kotlin comprehension, completion and code generation.
However, it would be interesting to further explore the capabilities of the classifier-based approach.

For instance, it is interesting how the prompt choice affects the model performance.
We hypothesize that different prompt choice can yield datasets which are more suitable for improving particular areas of model performance (\textit{e.g.}, knowledge of different packages, ability to generate high-quality tests, etc.). 

\subsubsection{KExercises}
The results on the KExercises dataset show a large pass rate increase, from 35\% to 60\% (or ten to 16 points) on the models we fine-tuned.
This enormous improvement comes with an improvement in syntax error rate similar to the one for KStack-clean. 
However, the changes in the completion scores are inconclusive with an improvement for Deepseek-7B and deterioration for the other models.
This could signify that fine-tuning on KExercises fits the model for the code generation from prompt, and it would be interesting to perform an exhaustive ablation study to find the limitations of KExercises and other datasets of this kind.
Our anecdotal experience with different prompts (such as the one we showcase on the HuggingFace page)\footnote{CodeLlama-7B-Kexer: \href{https://huggingface.co/JetBrains/CodeLlama-7B-Kexer}{https://huggingface.co/JetBrains/CodeLlama-7B-Kexer}} shows that the fine-tuning yields a model that understands various kinds of prompts, \textit{e.g.}, raw docstring.

\section{Future work}\label{sec:future-work}

This work covers only the essential basics for the Kotlin learning pipeline, including data and evaluation. 
However, we believe that the Kotlin ecosystem can offer much more to the language modeling community --- additional data, tools, and benchmarks.

We see the following directions for future research: 

\begin{itemize}
    \item \textbf{Learning from tools}. Several papers demonstrate the potential to improve models using feedback from various tools, such as compilers~\cite{zan2024codes}. The Kotlin ecosystem offers a vast array of tools that can be used in the learning process, including the compiler, linters, and translators. Additionally, modern IDEs possess a robust suite of inspections to check for code correctness, which could be directly integrated into the learning process.
    
    \item \textbf{Synthetic data}. In this paper, we presented KStack, which is a comprehensive representation of open-source Kotlin code. Despite our plans to keep it up-to-date, we will still miss many cases of Kotlin usage due to its production-oriented nature — we believe that a significant part of Kotlin code is stored in private repositories. As a workaround, we want to focus on generating more synthetic and high-quality code to cover not only coding exercises but also more realistic production tasks.
    
    \item \textbf{More benchmarks}. While in this work we suggest the most essential benchmark --- HumanEval, it can hardly represent real-life Kotlin tasks. In recent years, there has been a new generation of issue-based benchmarks like SWE-Bench~\cite{jimenez2024swebench}. Producing a similar dataset consisting of real-world Kotlin projects and issues would greatly benefit the assessment of whether the current generation of models can be useful in the day-to-day development process.
    
\end{itemize}

We hope that producing more artefacts for the Kotlin learning will not only help improve the quality of Kotlin code generation, but also provide researchers more tools and ideas for other languages.
\section{Conclusion}\label{sec:conclusion}

In this work, we aimed to achieve two primary goals: first, to produce a set of artifacts for the further development of Kotlin language models, and second, to abstract a set of techniques that aid in the development of language models for low-resource languages. 
We believe that the approach we used in developing the Kotlin ML Pack can be replicated for any other programming language, thereby enhancing current open-source language models for them.    
\pagebreak

\bibliographystyle{IEEEtran}
\balance
\bibliography{IEEEabrv,paper}

\end{document}